\documentclass[a4paper,11pt]{article}
\pdfoutput=1

\usepackage{jinstpub}
\usepackage{siunitx}
\usepackage{xcolor}
\title{Probing the angular and polarization reconstruction of the ARIANNA detector at the South Pole}

\author[a]{A.~Anker}
\author[a]{S.~W. Barwick}
\author[b]{H. Bernhoff}
\author[c,d]{D.~Z. Besson}
\author[e]{N. Bingefors}
\author[f,g]{D. Garc\'ia-Fern\'andez}
\author[a,1]{G. Gaswint \note{Corresponding author}}
\author[a, e]{C. Glaser}
\author[e]{A. Hallgren}
\author[h]{J.~C. Hanson}
\author[i]{S.~R. Klein}
\author[j]{S.~A. Kleinfelder}
\author[a,g]{R. Lahmann}
\author[c]{U. Latif}
\author[f, g]{Z.~S. Meyers}
\author[k]{J. Nam}
\author[c]{A. Novikov}
\author[f,g]{A. Nelles}
\author[a]{M.~P. Paul}
\author[a]{C. Persichilli}
\author[f,g]{I. Plaisier}
\author[a,l]{J. Tatar}
\author[k]{S.-H. Wang}
\author[f, g]{C. Welling}

\affiliation[a]{Department of Physics and Astronomy, University of California, Irvine, CA 92697, USA}
\affiliation[b]{Uppsala University Department of Engineering Sciences, Division of Electricity, Uppsala, SE-752 37 Sweden}
\affiliation[c]{Department of Physics and Astronomy, University of Kansas, Lawrence, KS 66045, USA}
\affiliation[d]{National Research Nuclear University MEPhI (Moscow Engineering Physics Institute), Moscow 115409, Russia}
\affiliation[e]{Uppsala University Department of Physics and Astronomy, Uppsala, SE-752
37, Sweden}
\affiliation[f]{DESY, 15738 Zeuthen, Germany}
\affiliation[g]{ECAP, Friedrich-Alexander-Universit\"at Erlangen-N\"urnberg, 91058 Erlangen, Germany}
\affiliation[h]{Whittier College Department of Physics, Whittier, CA 90602, USA}
\affiliation[i]{Lawrence Berkeley National Laboratory, Berkeley, CA 94720, USA}
\affiliation[j]{Department of Electrical Engineering and Computer Science, University of California, Irvine, CA 92697, USA}
\affiliation[k]{Department of Physics and Leung Center for Cosmology and Particle Astrophysics, National Taiwan University, Taipei 10617, Taiwan}
\affiliation[l]{Research Cyberinfrastructure Center, University of California, Irvine, CA 92697 USA}

\emailAdd{ggaswint@uci.edu, sbarwick@uci.edu, christian.glaser@physics.uu.se}

\abstract{The sources of ultra-high energy (UHE) cosmic rays, which can have energies up to \SI{e20}{eV}, remain a mystery. UHE neutrinos may provide important clues to understanding the nature of cosmic-ray sources. ARIANNA aims to detect UHE neutrinos via radio (Askaryan) emission from particle showers when a neutrino interacts with ice, which is an efficient method for neutrinos with energies between \SI{e16}{eV} and \SI{e20}{eV}. The ARIANNA radio detectors are located in Antarctic ice just beneath the surface. Neutrino observation requires that radio pulses propagate to the antennas at the surface with minimum distortion by the ice and firn medium. Using the residual hole from the South Pole Ice Core Project, radio pulses were emitted from a transmitter located up to \SI{1.7}{km} below the snow surface. By measuring these signals with an ARIANNA surface station, the angular and polarization reconstruction abilities are quantified, which are required to measure the direction of the neutrino. After deconvolving the raw signals for the detector response and attenuation from propagation through the ice, the signal pulses show no significant distortion and agree with a reference measurement of the emitter made in an anechoic chamber. Furthermore, the signal pulses reveal no significant birefringence for our tested geometry of mostly vertical ice propagation. The origin of the transmitted radio pulse was measured with an angular resolution of \SI{0.37}{\degree} indicating that the neutrino direction can be determined with good precision if the polarization of the radio-pulse can be well determined. In the present study we obtained a resolution of the polarization vector of \SI{2.7}{\degree}. Neither measurement show a significant offset relative to expectation.

}

\collaboration{ARIANNA collaboration}

\begin{document}
\maketitle

\section{Introduction}
\label{sec:intro}

Ultra-high-energy (UHE) neutrino detection probes the universe at energy scales beyond the reach of photons, giving astrophysics unique insights in the observation and location of extreme astrophysical sources. These extreme sources can shed light on outstanding questions in astrophysics, in particular what the sources of UHE cosmic rays are \cite{doi:10.1146/annurev.aa.22.090184.002233, PhysRevLett.117.241101}. This is because UHE cosmic rays at or near the source can interact hadronically with the surrounding media, or with photons of the cosmic microwave background, ultimately producing astrophysical or cosmogenic neutrinos, respectively, from charged pion decay \cite{PhysRevD.59.023002,doi:10.1146/annurev-astro-081710-102620}. Using charged cosmic rays such as protons and heavier nuclei to find sources instead of neutrinos is problematic as they are deflected by galactic, and extra-galactic magnetic fields, scrambling their source direction. Neutrinos, however, are electrically neutral and therefore travel in straight paths. This makes neutrino detection an excellent candidate for identifying the sources of UHE cosmic rays. In order to find these sources, it is crucial for a neutrino detector to be able to reconstruct the neutrino direction with excellent precision \cite{DecadalWhitePaper,DecadalWhitePaper2}.

The aim of the ARIANNA detector \cite{BARWICK2014} is to search for these sources of UHE neutrinos. ARIANNA complements IceCube \cite{PhysRevLett.117.241101} in extending the reach of neutrino detection to energies greater than \SI{e17}{eV}. In this energy range, the most effective means of measurement is through radio \textit{Askaryan} radiation \cite{1965JETP...21..658A,PhysRevLett.99.171101} created when UHE neutrinos interact within a dielectric medium. Ice is a dense and relatively transparent dielectric medium for radio waves due to an attenuation length on the order of a kilometer \cite{BARWICK2005}. Naturally, the large sheets of ice found in (Ant)artic regions serve as an excellent location for the ARIANNA detector. ARIANNA deploys autonomous detector stations with radio antennas placed just beneath the surface. The \SI{1}{km} scale of the station spacing means that each station is an independent neutrino detector, thus maximizing the overall sensitivity. Currently, ARIANNA has nine detector stations in Moore's Bay, Antarctica, and two additional detector stations located at the South Pole.

To reconstruct the direction of a neutrino, ARIANNA needs to be able to measure the incoming signal direction and polarization as well as the viewing angle, i.e., the angle at which the particle shower is observed with respect to the Cherenkov angle \cite{GlaserICRC2019}. Since the density of Antarctic ice varies near the surface 'firn', radio signals bend as they propagate up towards the ARIANNA surface detector. Thus, one needs to be able to model accurately how the ice affects the radio signal during propagation. This paper will focus on ARIANNA's ability to reconstruct the angular direction and polarization from a source deep in the ice. These reconstruction capabilities are also an important ingredient in the energy reconstruction \cite{GlaserICRC2019}.

The angular and polarization reconstruction accuracy of the ARIANNA station has been studied previously for the stations on the Ross Ice Shelf \cite{Glaciology}. Calibration transmitter antennas buried slightly below the ice surface emitted radio pulses toward the bottom water-ice interface at Moore's Bay. The absolute measurement of the arrival direction of the reflected signal agreed with expectation to within 1 degree or better \cite{CoreyICRC2015}. Reflected signals also demonstrated that polarization of the electric field was preserved during propagation and reflection \cite{Glaciology}. Though these studies were encouraging, they were mostly confined to nearly vertical propagation. The new data presented in this paper extends those prior studies to include a range of more representative propagation directions for neutrino-induced radio signals.

Another test of the angular and polarization reconstruction was performed with ARIANNA by observing cosmic rays \cite{NellesICRC2019}. Cosmic ray interactions in the atmosphere generate radio pulses, which are well-understood (e.g.~\cite{HuegeRewiev2016,GlaserErad2016}). Hence, such cosmic rays act as an in-situ calibration source. These signals are more representative of neutrinos than the previous study in that the signal-to-noise ratio and frequency-content are neutrino-like. The pulse forms are very similar; air showers and in-ice particle showers both produce short bipolar pulses with frequencies of order $\mathcal{O}(\SI{100}{MHz})$. Those ARIANNA stations configured with upward-facing LPDAs reconstruct the polarization and direction of incoming radio pulses over a much broader range of incoming angles and physical conditions than the previous study (see also \cite{GlaserICRC2019, NellesICRC2019}). A newly developed forward folding technique was used to reconstruct the 3 dimensional electric-field pulse from the measured voltages \cite{NuRadioReco}. The reconstructed polarization shows a resolution of \SI{7.06}{\degree} around the theoretical expectation \cite{GlaserICRC2019, NellesICRC2019}. That study uses signals from the air, which excludes any effects from the ice. ARIANNA searches for neutrinos coming from the ice, however, and thus a test of the effects of the ice is crucial.

For such a test, a radio pulser was lowered in a fluid-filled hole provided by the South Pole Ice Coring Experiment (SPICE) to a depth of \SI{1.7}{km} \cite{SPICEcore}. The main focus of this paper will be to test systematic uncertainties of the direction and polarization reconstruction from ice propagation and detector calibration using this SPICE data set, and with this, determine the capabilities of the ARIANNA detector to resolve the direction of an incoming neutrino.

\section{Measurement Setup}
Data used for this measurement were collected by an ARIANNA South Pole station, which will be refereed to as station 51 throughout this paper, during the last week of December, 2018. The signal transmitter (IDL-1 pulser \cite{SPICEpulser}) was connected to a bicone antenna which was lowered to a depth of \SI{1.7}{km} inside the SPICE hole and was vertically-oriented (to match the form-factor of the SPICE hole) \cite{SPICEcore}. The IDL-1 pulser broadcasts short duration radio frequency pulses through the bicone antenna with a repetition rate of \SI{1}{Hz}, which is then detected by the ARIANNA station.  Several thousand pulser events were directly transferred over the Iridium satellite network for offline analysis of the angular and polarization reconstruction capabilities of ARIANNA.

\begin{figure}
    \centering
    \includegraphics[width=0.9\textwidth]{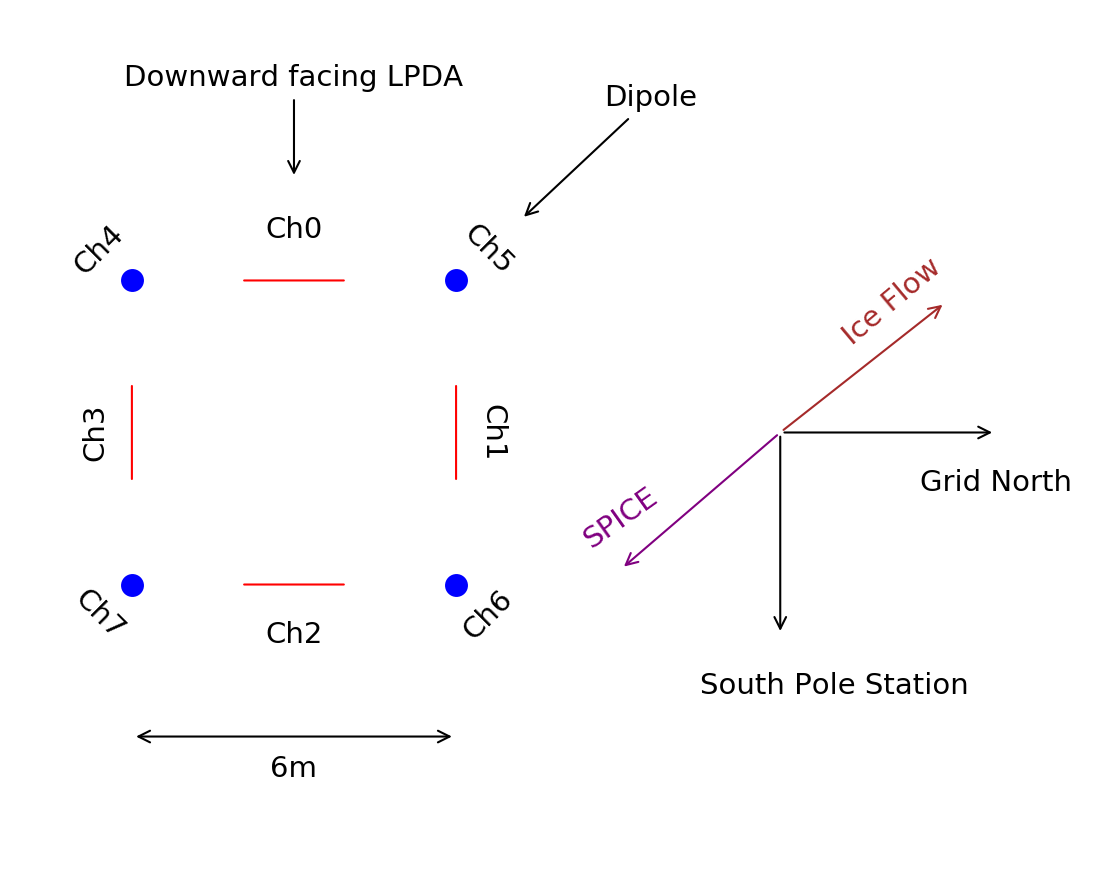}
    \caption{Overhead view of station 51 layout with direction to SPICE hole. Channels 0-3 are the 4 downward facing LPDAs buried beneath the surface, and channels 4-7 are the 4 vertically-oriented dipole antennas also burried beneath the surface. The angle between the ice flow direction (see \ref{sec:bire}) and signals coming from the SPICE borehole is \SI{1.4}{\degree}. }
    \label{fig:stn51}
\end{figure}

\subsection{Geometry and ARIANNA station at South Pole}
Station 51 is located roughly \SI{1}{km} from South Pole Station and \SI{0.65}{km} from the SPICE hole. Although there are no direct measurements, it is nevertheless plausible that the SPICE hole may be tilted by \SI{1}{\degree} which translates to a systematic uncertainty in the relative position of the emitter with respect to the detector station, which in turn results in a systematic uncertainty in the predicted signal arrival direction.

Station 51 is equipped with 8 antennas \cite{COSPAR2018}. Four of these are down-facing (nose pointing at \SI{180}{\degree} from zenith) Create CLP5130-2N log-periodic dipole arrays (LPDAs) oriented in a square pattern with \SI{6}{m} sides. The longest ($\frac{\lambda}{2}$) dipole is \SI{1.45}{m}, and the length ratio of adjacent dipoles is \SI{0.83}{}. The boom holding the dipoles in place has a length of \SI{1.385}{m}, and the cable feed point is at the shortest dipole ~\cite{LPDA}. The top of the LPDAs were located during deployment at a depth of \SI{~0.5}{m} into the firn and is subject to snow accumulation. Additionally, there are four vertically-oriented Kansas University bicone antennas located at the corners of the square. The bicone antennas are \SI{0.52}{m} in length, with the cable feed point at the top. The top of the bicone antennas were also located at a depth of \SI{0.5}{m}. The LPDAs provide two orthogonal Hpol (polarization parallel to the surface) measurements, whereas the bicones measure the Vpol component (vertical polarization). We provide a layout of station 51 in Fig.~\ref{fig:stn51}.

\subsection{Characteristics of the signal transmitter and ice propagation}
The characteristics of the signal transmitter were tested in an anechoic chamber and combined with simulations of the known ice effects on signal propagation from transmitter to receiver in the SPICE run \cite{Barwick2018,NuRadioMC}.

\begin{figure}
    \centering
    \includegraphics[width=0.5\textwidth]{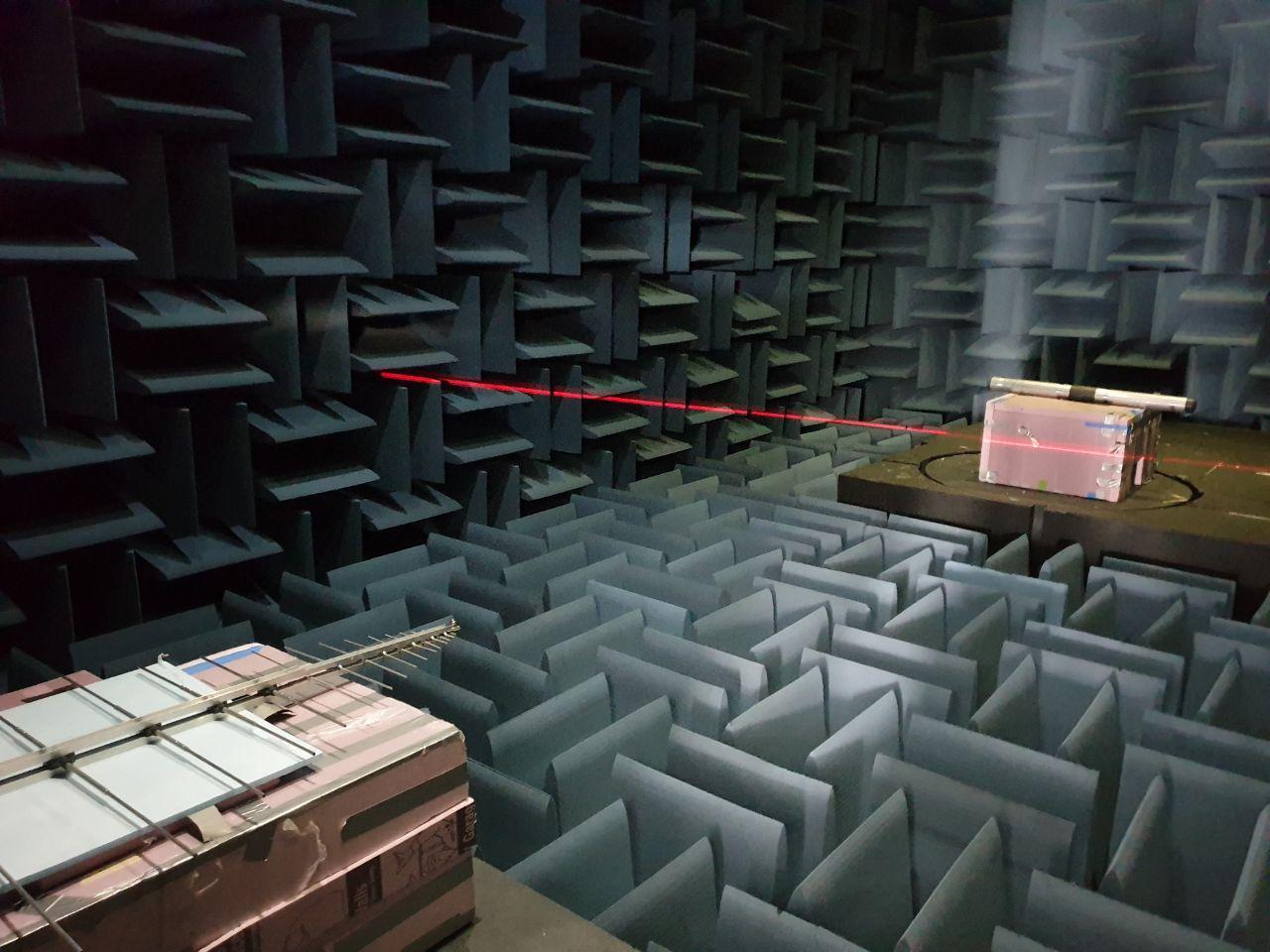}
    \caption{Photo of the anechoic chamber experimental setup. The transmitting bicone antenna was rotated horizontally. The receiving LPDA antenna was orientated in two ways for every measurement. The first orientation being what is shown in the photo (tines laid horizontally), and the second orientation had the tines vertically oriented which was stabilized with foam bricks.}
    \label{fig:anechoicPic}
\end{figure}

\subsubsection{Anechoic chamber measurement of signal emitter}
\label{anechoicPol}
The IDL-1 pulser with the same bicone antenna used in the SPICE measurements, and an ARIANNA LPDA receiver were set up in an anechoic chamber to make a prediction for the polarization expected in the SPICE data. The separation distance between receiver and transmitter in the anechoic chamber was \SI{3}{m} and the data recorded with a \SI{5}{GSa/s} sampling oscilloscope. The anechoic chamber has dimensions \SI{11.58}{m} x \SI{7.29}{m} x \SI{7.36}{m} \cite{KUanechoicDimension}. Fig.~\ref{fig:anechoicPic} shows an image of the measurement setup.

To test the polarization calibration as a function of launch angle, the transmitting antenna was rotated between \SI{0}{\degree} and \SI{90}{\degree} in the horizontal plane while the receiving LPDA was pointing towards the dipole antenna and orientated either at \SI{0}{\degree} (tines parallel to ground) or \SI{90}{\degree} (tines perpendicular to ground). This allowed us to capture the two polarization components of the emitted electric field for a range of launch angles. For each setup, 10 individual measurements were recorded.

The anechoic chamber data are processed in the same way as the SPICE data, as described in the next section. To account for the difference in dielectric environments, after reconstructing the anechoic chamber electric-field, the frequency content is shifted from an in-air medium to an in-ice medium by dividing by the index of refraction of deep ice ($n = 1.78$). Shifting the frequencies by $\frac{1}{n}$ serves as a first-order approximation since the antenna is wavelength-resonant; to convert from a wavelength to a frequency in a different medium, a factor of $n$ must be applied \cite{NuRadioReco}. After performing this frequency correction, a rectangular band pass filter between \SI{80}{MHz} to \SI{300}{MHz} is applied in order to remove unwanted noise. Lastly, the signals are up-sampled to \SI{50}{GHz} for better time resolution.

\begin{figure}
    \centering
    \includegraphics[width=0.6\textwidth]{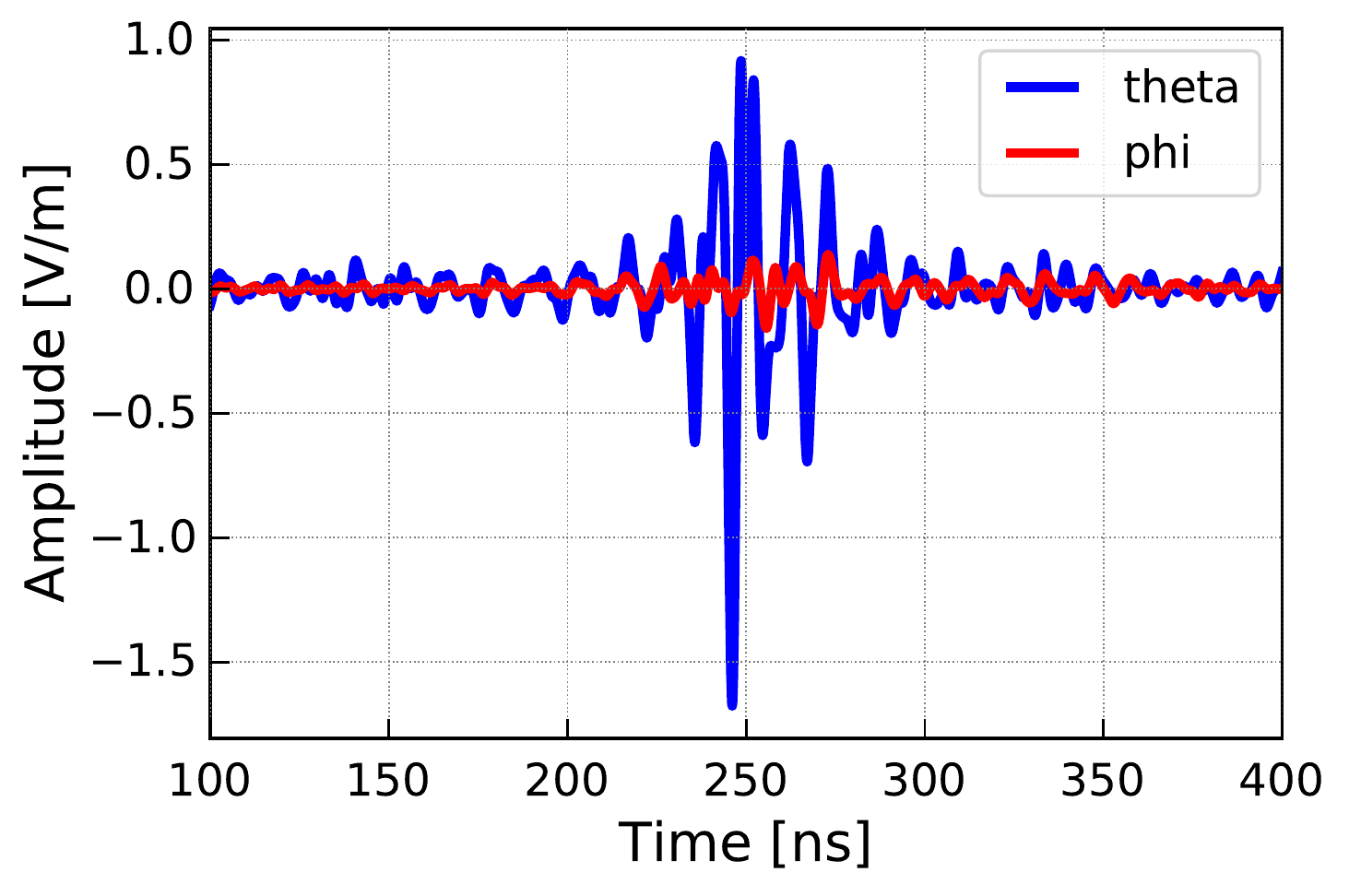}
    \caption{Observed electric-field from the IDL-1 pulser at an angle of \SI{30}{\degree} off boresight and captured inside an anechoic chamber. The LPDA antenna acts as receiver and a bicone antenna was used as the emitter. The LPDA antenna response was factored out of the voltage traces to obtain the electric-field. The theta polarization corresponds to the polarization along the axis of the dipole, while phi polarization is the cross polarized component.}
    \label{fig:transmittedEField}
\end{figure}

\begin{figure}[t]
    \centering
    \includegraphics[width=0.5\textwidth]{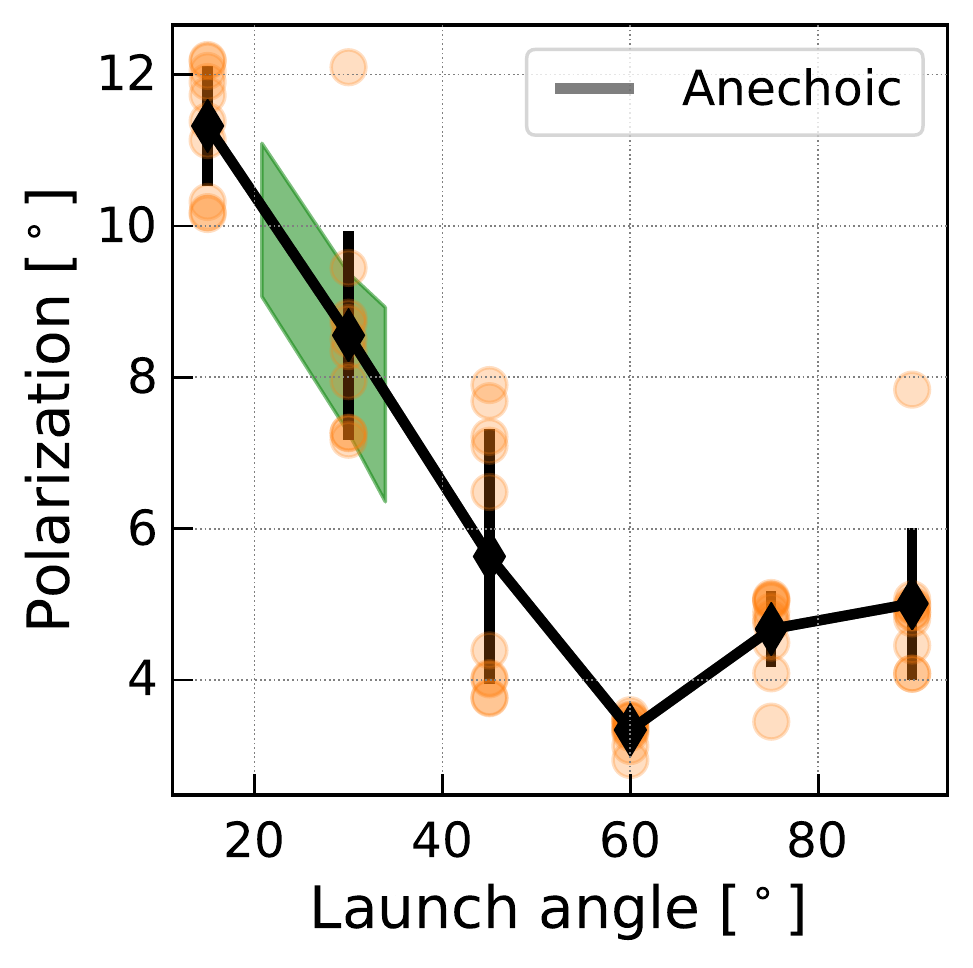}
    \caption{Expected polarization angle of the received electric field as a function of transmitted angle with respect to the main symmetry axis of the dipole. The orange data points are the polarization angle's found from using a single pair of voltage measurements. The black data points are the averages of the orange data, with the error bars being the standard deviation of the orange data points. Green shows the relevant transmitted angles for the SPICE data with a 1$\sigma$ spread based off of the 16\% and 84\% quantile.}
    \label{fig:expPol}
\end{figure}

An example of a transmitted pulse taken from the anechoic chamber, following these corrections, is shown in Fig.~\ref{fig:transmittedEField} This example pulse was emitted at an angle of \SI{60}{\degree} off the direction of maximal gain (a typical geometry in the SPICE data). The electric field is mainly theta-polarized (polarized along the main symmetry axis of the dipole). This serves as a baseline signal for the ARIANNA polarization reconstruction. Defining the polarization as the angle between the energy fluence of the theta and phi polarization (see Sec.~\ref{sec:pol_reco} for more details), the signal polarization measurement derived from the average of the calculated electric-field magnitudes from the 10 measured voltages for each polarization captured for a given geometry is shown in Fig.~\ref{fig:expPol}, which shows that the polarization reconstruction relevant for the launch angles in the SPICE experiment are between \SI{8}{\degree} to \SI{10}{\degree} (highlighted by the green band).

We note that this calibration signal is more difficult to reconstruct than a neutrino-induced signal for two reasons:

\begin{enumerate}
    \item \textbf{Polarization:} The anechoic signal is almost entirely polarized in the theta direction. Therefore, the noise in the weaker phi component will have a large influence on the polarization reconstruction. For neutrinos, the signals are expected to have comparable theta and phi content.
    \item \textbf{Pulse shape:} The phi component of the anechoic chamber signals have an extended pulse form, and with different frequency content compared to the theta component. A minimum integration window of \SI{70}{ns} is necessary to sufficiently capture both components. Neutrinos will have signals with polarization projections equal in both length and frequency and differing only in amplitude between the two components. Therefore the polarization reconstruction will not strongly depend on the integration window and frequency cut.
\end{enumerate}
With this in mind, this analysis can be considered a lower bound on the polarization reconstruction capabilities of neutrino signals, which should give cleaner signals. 

\subsubsection{Calculation of incoming signal direction}
\label{iceProp}
\begin{figure}
    \centering
    \includegraphics[width=0.3\textwidth]{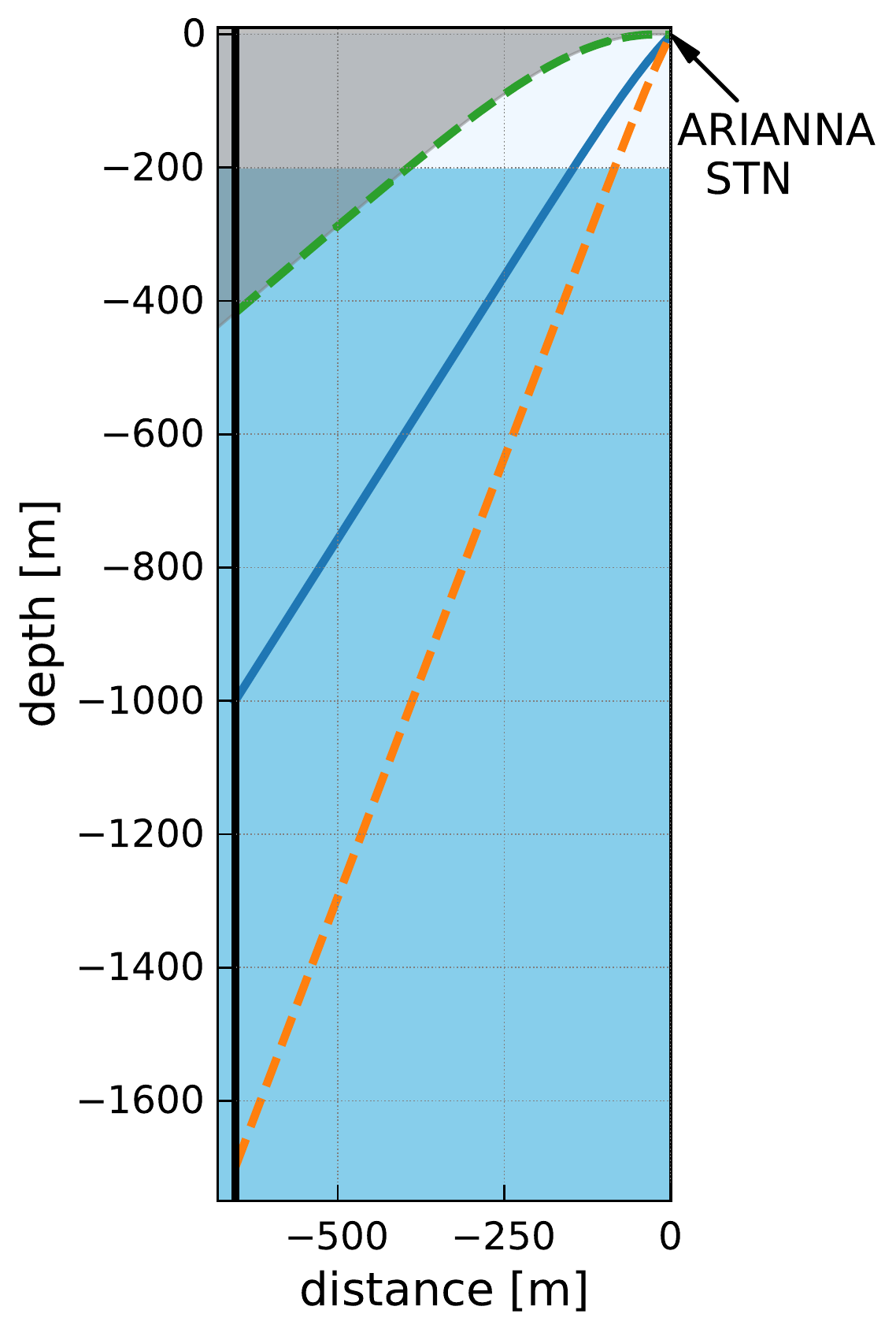}
    \includegraphics[width=0.45\textwidth]{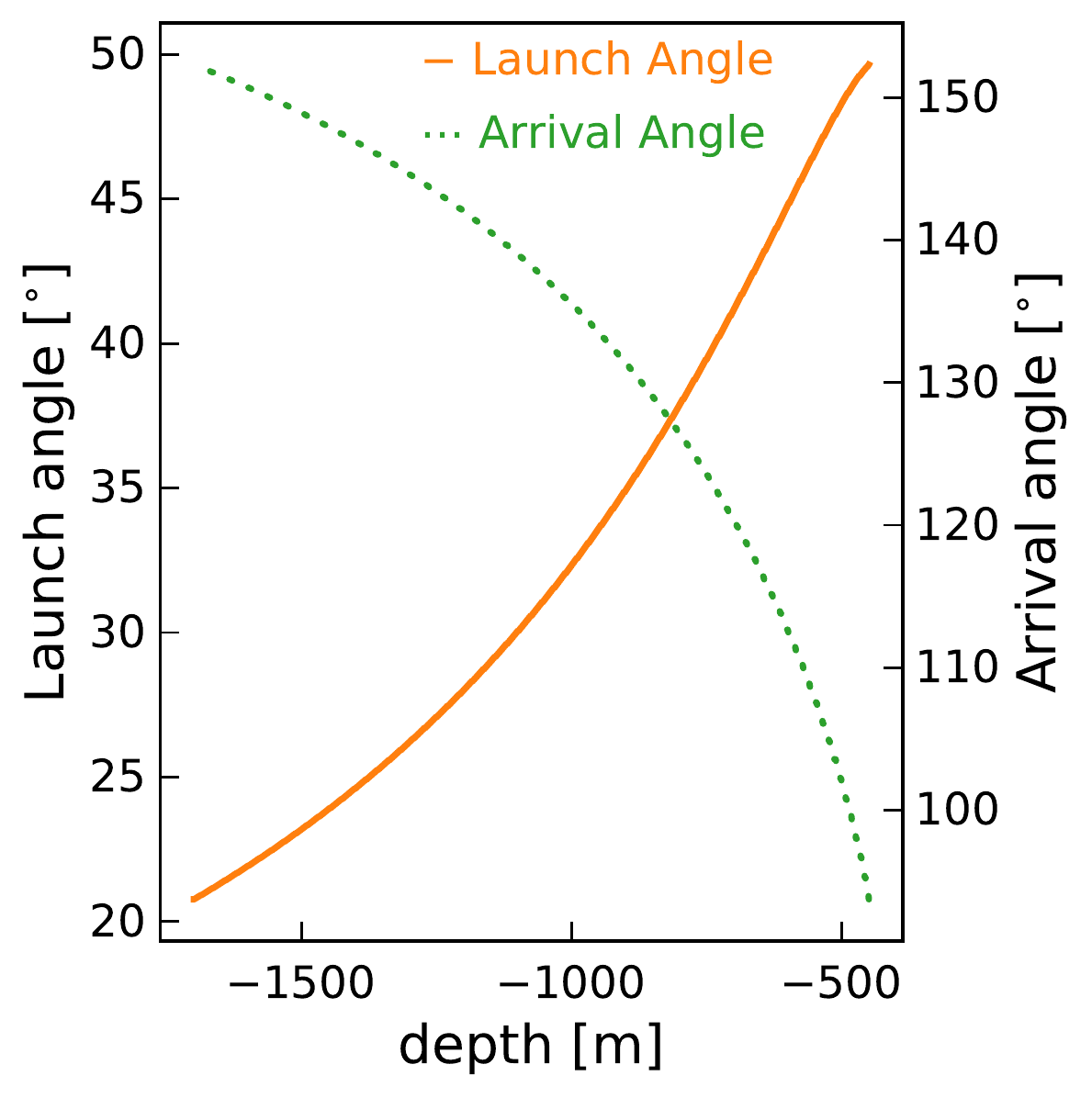}
    \caption{(Left) Ray tracing solution from a transmitter at depths \SI{418}{m}, \SI{1}{km} and \SI{1.7}{km} to the South Pole ARIANNA station 51 calculated with the NuRadioMC code \cite{NuRadioMC}. Light blue shaded region above \SI{200}{m} is the firn layer, over which the ice approaches (within 2\%) its nominal density. The grayed  area is the shadow zone, from which classical propagation to the station is forbidden. The vertical black line on the y-axis represents the SPICE borehole. (Right) Expected arrival zenith angle and expected launch zenith angle as a function of transmitter depth. \SI{180}{\degree} corresponds to the nadir.}
    \label{fig:rays}
\end{figure}

Testing the ARIANNA angular reconstruction capabilities requires an accurate model of how ice affects propagation. The density, and therefore the index-of-refraction, changes in the upper \SI{200}{m} of the South Pole ice sheet from $n = 1.78$ for deep ice to about $n = 1.35$ at the surface. As a consequence, radio signals do not propagate rectilinearly but are refracted as illustrated in Fig.~\ref{fig:rays} \cite{NuRadioMC}. We use an exponential index-of-refraction ($n$) vs depth ($z$) profile which provides a good description of $n(z)$ data that was derived from density measurements \cite{Barwick2018,indexVsDepth}. The gray shading indicates the range of positions of the transmitter that permit no classical propagation solutions, which is termed the 'shadow zone'. Signals in the shadow zone, bend back into the ice before reaching the ARIANNA detector. However, signals can be seen in the shadow zone through horizontal propagation, likely due to deviations from a purely exponential density profile \cite{Barwick2018, RobertICRC2019}. These effects are not discussed in this paper. 

Three representative allowed solutions are displayed in Fig.~\ref{fig:rays} corresponding to pulser depths of \SI{418}{m}, \SI{1}{km} and \SI{1.7}{km}. The right panel gives the expected arrival and launch zenith angle (measured with respect to \SI{0}{degrees} zenith) at the ARIANNA station as a function of transmitter depth.

\section{Processing of data taken in the field}
This section describes how data from radio
pulses emitted by the antenna in the SPICE hole are processed and which additional calibration steps had to be performed to reduce systematic uncertainties.

\subsection{Main processing steps}
Four main processing steps are applied to all events from the measurement campaign from the SPICE borehole:

\begin{enumerate}
    \item \textbf{Initial quality cut:} The readout electronics of the ARIANNA station become non-linear when the signal amplitude exceeds \SI{600}{mV}. During the 2018 SPICE core run, events in the linear regime occur at depths greater than \SI{800}{m} and only these events are retained for analysis. 
    \item \textbf{Band pass filter:} To reduce out-of-band noise, the frequency content of the events is restricted with a rectangular band pass filter to between \SI{80}{MHz} (set by the frequency threshold of the receivers) and \SI{300}{MHz}. This cut also reduces the influence of noise on the time correlations of the signal pulses, improving the accuracy of the direction reconstruction.
    \item \textbf{Deconvolution of signal chain:} To properly compare measured data from different channels, the amplifier response is deconvolved along with time delays from cables and electronics (as measured in the lab).
    \item \textbf{Upsampling:} The traces are up-sampled from \SI{1}{GHz} to \SI{50}{GHz}, using the Fourier method provided by resample from the scipy package in python \cite{scipy}, to improve the timing resolution. This allows us to correlate the signals to \SI{0.02}{ns} accuracy.
\end{enumerate}

\subsection{In-situ calibration of cable delays}

The cables in ARIANNA station 51 were measured with a precision of \SI{0.5}{ns}. We can use the data itself to improve this calibration to about \SI{0.1}{ns} by the following procedure. For each data point we calculate the expected propagation time from the emitter to each receiving antenna using the signal propagation (ray tracing) module of NuRadioMC \cite{NuRadioMC}. Also for each data point, we calculate the time differences between the signal pulses received in the antennas (separately for the LPDAs and the dipoles) by cross-correlating the signal pulses against the signal pulse of one reference channel, chosen arbitrarily as channel 3 for the LPDAs and channel 6 for the dipoles. Knowing the source location and the ray trajectories, we then subtract the expected time delays from signal propagation from the measured time delays. 
We find largely constant time offsets that are compatible within the experimental uncertainties of the station calibration. 
The distributions are approximately Gaussian with offsets of up to \SI{1}{ns} between channels and standard deviation up to $\sim$\SI{0.1}{ns}. The variation is much smaller than the mean offset, though there is a slight depth dependence that was not consistent between channels. The mean of the distribution is assumed to be associated with cable delays or other delays along the signal chain.
These time offsets ($\Delta T$) are presented in Tab.~\ref{table:timeOffsets} and are added to the cable delays when deconvolving the signal chain (step three above). We note that this procedure does not necessarily center the mean of the expected arrival direction (using all antennas) to the predicted arrival direction since we have used a single reference channel for our calculation of $\Delta T$. 

\begin{table}[ht]
\centering
\begin{tabular}{c  c  c  c} 
 \hline \hline
 Channel & Reference Channel & Mean [ns] & STD [ns] \\ [0.5ex] 
 \hline
 0 & 3 & -1.34 & 0.12 \\
 1 & 3 & -0.70 & 0.09 \\
 2 & 3 & -0.16 & 0.07 \\
 3 & 3 & 0.0 & 0.0 \\
 4 & 6 & 0.11 & 0.12 \\
 5 & 6 & -0.07 & 0.10 \\
 6 & 6 & 0.0 & 0.0 \\
 7 & 6 & -0.99 & 0.09 \\ [1ex] 
 \hline \hline
\end{tabular}
\caption{Time differences between channels after deconvolution of the hardware response and subtracting the expected time delays for each individual channel. First 4 rows use channel 3 (LPDA) as the reference channel, whereas the last four rows use channel 6 (dipole) as a reference channel. The mean of the time delay offsets from zero can be associated with uncertainties in cable delays.}
\label{table:timeOffsets}
\end{table}

\section{Direction reconstruction and angular resolution}
\label{sec:direction}

The direction reconstruction capabilities of ARIANNA were previously reported at the 2019 ICRC \cite{GaswintICRC2019}. Here, we report on the same study but with a larger data set. The analysis has been improved by accepting a larger bandwidth for the angular reconstruction, and improving the precision of the time delays between channels. The overall conclusion, however, is unchanged.

The incoming direction of a triggered event can be reconstructed through the timing delays between antennas. The NuRadioReco framework \cite{NuRadioReco} is used to reconstruct the incoming direction of a triggered event in the ARIANNA detector. The particular algorithm used is called the \emph{cross correlation method} as it uses the time differences between two parallel pairs of antennas (found through correlating the two signals together) to determine the signal arrival direction. (See \cite{NuRadioReco} for details of the reconstruction algorithm.) Correlating two signals together is typically improved with a filter in the time domain, and thus for the angular reconstruction, a Hanning window with a rise time of \SI{20}{ns} and a width of \SI{50}{ns}, and for which the filter is centered around the pulse maximum is applied. This aids the reconstruction by reducing the influence of noise and by removing after-pulses and other artifacts that could lead to spurious correlations of the trace not associated with the main signal.

\begin{figure}
\centering
\includegraphics[width=1.0\textwidth]{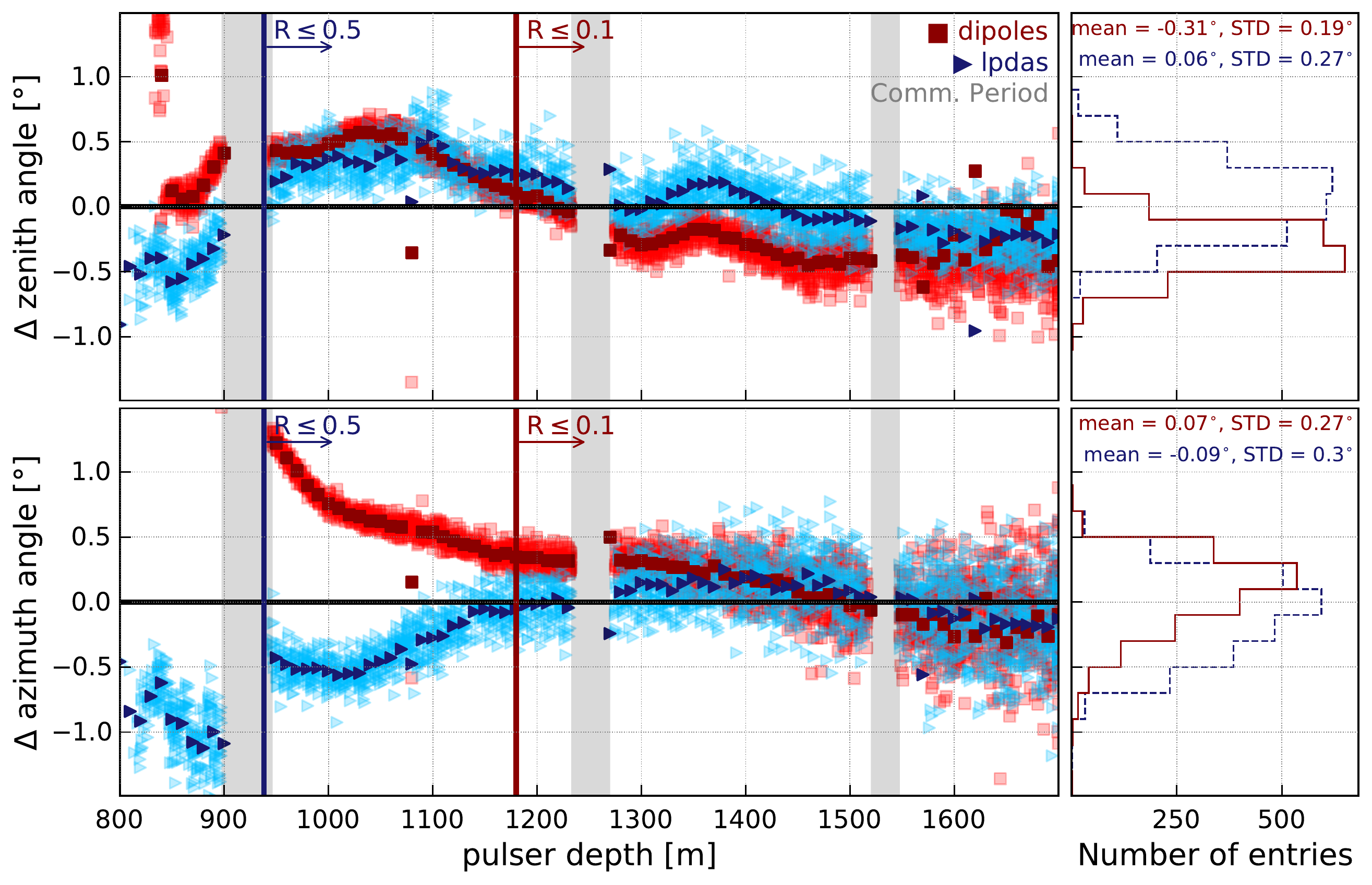}
\caption{Reconstructed arrival direction minus expected arrival direction. Left plots show the depth dependence; histogram projections are shown on the right. This data is corrected for the time differences between channels shown in Table \ref{table:timeOffsets}. The expected arrival direction is found using the NuRadioMC ray tracer while the reconstructed arrival direction is found through the cross correlation method. Light blue triangles show the residuals using the four LPDAs along with a \SI{10}{m} average shown in a darker blue color. Red squares show the residuals using the four dipoles along with a \SI{10}{m} average shown in a darker red color. Each average has roughly 30 events. The red vertical line corresponds to a reflection coefficient of 0.1, while the blue vertical line corresponds to a reflection coefficient of 0.5. The gray shaded area indicates the periods where station 51 was in communication mode and thus not taking data. The data in the projected histograms present the residuals on an event-by-event basis (i.e. without the averaging). Blue dashed is used for LPDAs, and red is used for dipoles. For the LPDAs all data-points with $R\leq0.5$ are included and for the dipoles all data-points with $R\leq0.1$ are included (see text for details). The mean and standard deviation is reported in the upper right corner of the histograms.}
\label{fig:angularReco}
\end{figure}

We measure the arrival direction using the four LPDA waveforms; an independent measurement from the four dipole antennas provides a cross-check. In Fig.~\ref{fig:angularReco}, we present the reconstructed signal arrival directions, relative to prediction, as a function of the emitter depth (cf.~Fig.~\ref{fig:rays}). The full range of SPICE data is included in Fig.~\ref{fig:angularReco} along with an average of \SI{10}{m} depths of the reconstructed angular directions. Each point in the averaged data has roughly 30 events. For the LPDAs we apply an additional cut on the data where the reflection coefficient of the firn-air boundary is $50\%$ or less which corresponds to a depth of \SI{938}{m}. This extra cut is applied to minimize the influence of interference between reflected and direct signals arriving at the receiving LPDAs. We find a resolution in azimuth and zenith to better than \SI{0.3}{\degree} centered around \SI{-0.1}{\degree} and \SI{0.1}{\degree} respectively. The dipoles are equally sensitive to signals arriving from above and below and were buried just \SI{0.5}{m} below the surface and therefore the interference between reflected signals and direct signals is more pronounced. Thus, we apply a more stringent cut to the dipoles, requiring the reflection coefficient is $10\%$ or less which corresponds to a transmitter depth of \SI{1180}{m}. The dipoles give a resolution of \SI{0.2}{\degree} in zenith with a \SI{-0.3}{\degree} offset and resolution of \SI{0.3}{\degree} in azimuth centered around \SI{0.1}{\degree}. 

Another Askaryan based neutrino detector, ARA, has looked at reconstructing deep pulser events \cite{Allison_2016}. The ARA experiment uses birdcage dipoles for the vertical polarization and ferrite loaded quad-slot antennas for the horizontal polarization buried at depths between \SI{170}{m} and \SI{190}{m} which greatly reduces firn effects on the signal propagation. ARA reports an azimuthal resolution of \SI{1.3}{\degree} or better, with an offset of up to \SI{2.0}{\degree}. Without taking any firn effects into account, ARA reports sub-degree precision of at most \SI{0.4}{\degree} in zenith, albeit with a systematic offset of up to \SI{4.8}{\degree} \cite{Allison_2016}.

To estimate the resolution on the ARIANNA directional reconstruction, the 3D angular difference between the reconstructed and predicted arrival direction is calculated. For the LPDAs, ARIANNA achieves a directional resolution of \SI{0.37}{\degree} whereas for the dipoles ARIANNA achieves a resolution of \SI{0.43}{\degree}. If we do not apply a depth cut to remove reflections, but instead take all data from when the transmitter was at depths greater than \SI{800}{m} into consideration, then ARIANNA is able to achieve an angular resolution of \SI{0.41}{\degree} using the LPDAs and \SI{0.55}{\degree} using the dipoles. The measurement of the radio incoming direction is important for an accurate reconstruction of the vertex direction and the neutrino direction reconstruction (see \cite{GlaserICRC2019}).

The slight offset seen in the zenith reconstruction using the dipoles is due to the four dipoles recording slightly different pulse shapes. It was assumed that the 4 dipoles had the exact same antenna response. The offset suggests significant antenna to antenna variations which we speculate are due to the $\sim$\SI{50}{cm} proximity of the dipoles to the surface, with additional uncertainties associated with possible slight variations in orientations. 
Further investigations of the antenna-to-antenna response will hopefully mitigate any variations in the dipole zenith reconstruction.\footnote{We note that the proposed ARIANNA-200 detector \cite{ARIANNA200} aims to install the same mark dipoles at a depth of at least \SI{5}{m}, and that the antenna depth will naturally increase with time due to snow accumulation. Thus, we expect that antenna-to-antenna variations will be much smaller than observed here.}

There are some slight depth dependencies seen in 
Fig~\ref{fig:angularReco}, which may result from:
\begin{itemize}
    \item \textbf{Ice profile:} An uncertainty in the index-of-refraction profile used to predict the signal arrival direction from the depth of the emitter will affect the prediction of the zenith angle. However, the azimuthal angle would remain unaffected as ice cannot affect the angle orthogonal to signal propagation (under the assumption of a vertical index of refraction profile, without horizontal components). The residuals from LPDAs and dipoles are affected in the same way.  
    \item \textbf{Tilt of SPICE hole:} To predict the signal arrival direction we assume that the SPICE borehole is straight down, i.e., only the z-position of the emitter changes. However, the hole could have a tilt to up to \SI{1}{\degree} which would lead to a change in the zenith and/or azimuth prediction as a function of depth depending on the direction of the tilt. The residuals from LPDAs and dipoles are affected in the same way.  
    \item \textbf{Antenna position:} Uncertainties in the antenna position can lead to uncertainties in the directional reconstruction that are dependent on the signal arrival direction and therefore depth-dependent. In this case, the residuals from LPDAs and dipoles would be affected differently. 
    \item \textbf{Antenna response:} Differences in the antenna response within the separate sets of dipoles and LPDAs can lead to antenna-dependent and signal arrival direction-dependent time delays and pulse distortions. Although mechanical differences are unlikely to cause any significant difference, since the antennas are so shallow, the close vicinity to a boundary is likely to influence the antennas differently. This effect should be mostly visible at signal arrival directions for which the surface becomes reflective. Furthermore, the effect should be mostly visible in the dipoles because they are equally sensitive to upward and downward coming signals, whereas the LPDAs have a reduced gain for signals entering the (downward-facing) LPDAs from above. 
\end{itemize}
We observe the strongest deviations from the prediction for dipoles at signal arrival directions for which the surface become reflective. We attribute this to uncertainties in the antenna response. The scatter of the LPDA reconstruction is also larger when the surface becomes reflective, although the effect is less pronounced than for the dipoles, consistent with this hypothesis. There is no consistent depth dependence between the LPDA and dipole reconstruction which disfavors a dominant influence of the ice profile or tilt of the SPICE hole. In particular, we can conclude that the ice is understood well enough to correct signal arrival directions for the ray bending to better than \SI{1}{\degree}, an important result for reconstruction of the neutrino direction. \footnote{We also note that the antenna-related uncertainties will improve in a future ARIANNA-200 detector \cite{ARIANNA200} by installing the antennas deeper into the ice to reduce interference between reflected and direct signals.}

\section{Measurement and interpretation of the signal polarization}
\label{sec:pol_reco}
In this section, the reconstruction of the polarization of the SPICE pulser signals is presented. 
To measure the polarization, ARIANNA needs to be able to measure the electric field using at least two perpendicular antennas. Using the two orthogonally oriented LPDAs, the framework NuRadioReco ~\cite{NuRadioReco} is used to reconstruct the electric field from the recorded voltage traces. The electric field is reconstructed by solving the following system of equations.

The electric field $\mathcal{E}^{\phi,\theta}$ relates to the voltage output $\mathcal{V}_i$ of an antenna $i$ in Fourier space as
\begin{equation}
    \begin{pmatrix} \mathcal{V}_1(f) \\ \mathcal{V}_2(f) \\ ...\\ \mathcal{V}_n(f)\end{pmatrix} = 
    \begin{pmatrix} \mathcal{H}_1^\theta (f)& \mathcal{H}_1^\phi (f)\\ \mathcal{H}_2^\theta (f) & \mathcal{H}_2^\phi (f)\\ ... \\ \mathcal{H}_n^\theta (f)& \mathcal{H}_n^\phi (f)\end{pmatrix} 
    \begin{pmatrix} \mathcal{E}^\theta(f) \\ \mathcal{E}^\phi(f)\end{pmatrix} \, ,
    \label{eq:H_full}
\end{equation}
where $\mathcal{H}_i^{\theta, \phi}$ represents the response of antenna $i$ to the $\phi$ and $\theta$ polarization of the electric field $\mathcal{E}^{\theta, \phi}$ arriving from a particular direction. The polarization states $\theta$ and $\phi$ are the two orthogonal vectors in spherical coordinates that are perpendicular to the signal propagation direction. For a horizontally propagating signal, $\vec{e}_\phi$ lies in the horizontal plane whereas $\vec{e}_\theta$ is oriented vertically.

The SPICE data was measured with 2 pairs of LPDA antennas with orthogonal polarization sensitivity. We apply a linear least square minimization to extract the electric field vector from the overdetermined system of equations. The anechoic chamber measurement was performed with just two orthogonal LPDAs which leads to an exact solution of Eq.~\eqref{eq:H_full}.  The polarization is then calculated from the electric fields via:
\begin{equation}
    P = \arctan{\frac{f_{\phi}}{f_{\theta}}}\label{eq:1}
\end{equation}
with,
\begin{equation}
    f_{\phi} = \sqrt{ \sum\limits_{t=t_m - \SI{35}{ns}}^{t_m+\SI{35}{ns}}{|E_{\phi}(t)^2|}} - f_{\phi,\mathrm{noise}}\label{eq:2}
\end{equation} 
where $f_{\phi}$ is the energy fluence for the ${\phi}$ component, $E_{\phi}$ is the electric field for the ${\phi}$ component, and $P$ is the polarization. The time $t_m$ is the position of the maximum of the Hilbert envelope of the (dominant) $\theta$ component of the electric field. The $\theta$ component of the energy fluence is defined analogously. The quantity $f_{\phi,\mathrm{noise}}$ is an estimate of the noise contribution which is calculated from a part of the recorded trace that does not contain signal. This definition is general and robust against different experimental configurations such as the anechoic chamber data vs.~the SPICE data, and can also be directly applied to a neutrino event. Because noise is subtracted, this definition is also largely independent of the exact choice of the integration window; this was confirmed by analyzing the data with different choices of integration windows.

\subsection{Polarization reconstruction and resolution}
The transmitting angles for the range of depths that was analyzed by ARIANNA in the SPICE data are between \SI{21}{\degree} and \SI{32}{\degree} (with respect to the vertical) and which is also highlighted in green in Fig.~\ref{fig:expPol}. These angles are determined through the ray-tracing solutions found using NuRadioMC as outlined in Sec.~\ref{iceProp}. The expected polarization angles for this depth-range are between \SI{8}{\degree} and \SI{10}{\degree}, see Sec.~\ref{anechoicPol}. Ice effects, including the bending of the signal, and the frequency-dependent ice attenuation are accounted for in this calculation. The ice attenuation used is the \emph{South Pole simple} model in NuRadioMC~\cite{NuRadioMC} and is derived from RICE data gathered in 2004 ~\cite{Barwick2018}. 

\begin{figure}[t]
\centering
\includegraphics[width=0.8\textwidth]{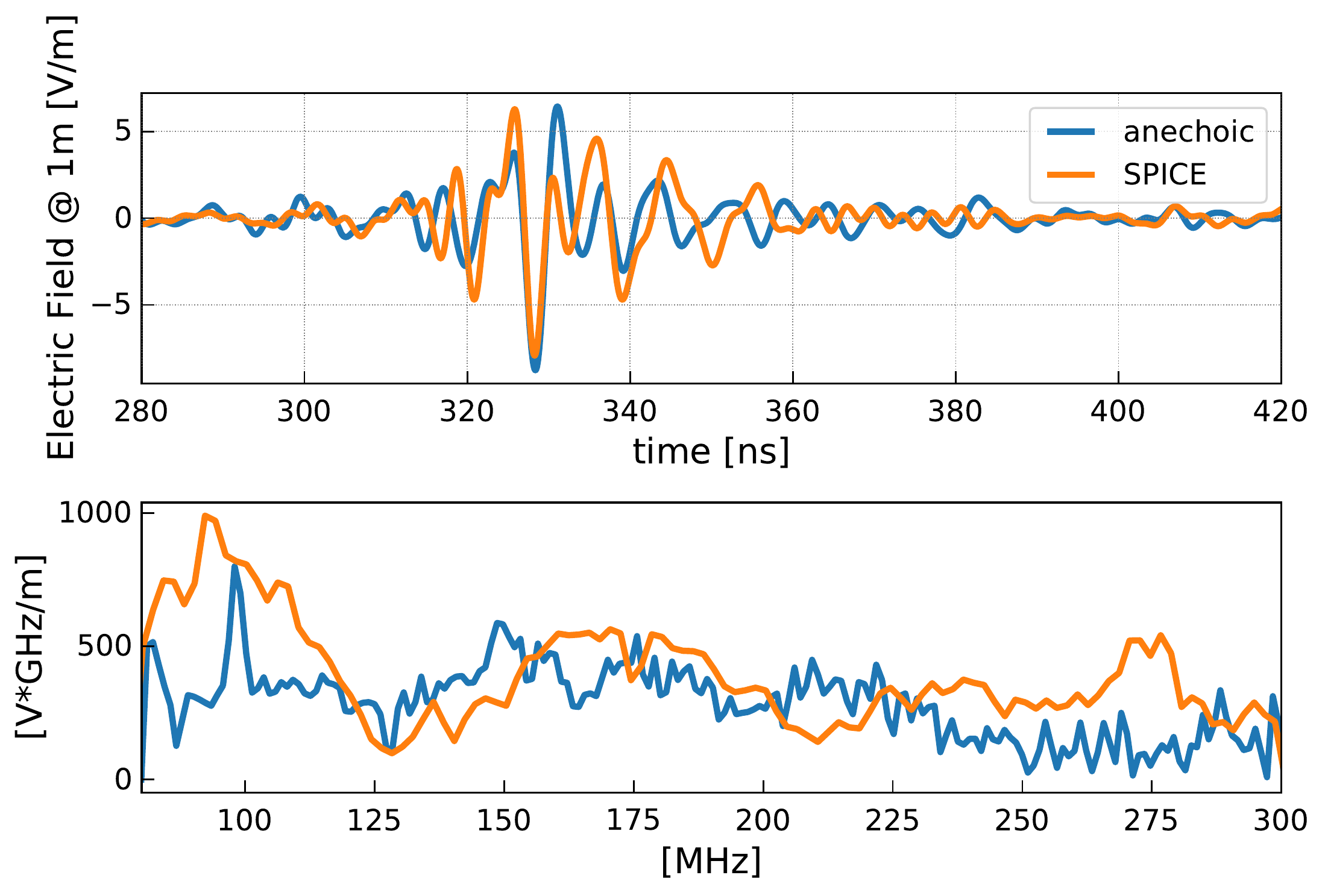}
\caption{\label{fig:efieldCompare} Overlays the reconstructed electric field from 2019 SPICE hole experiment (including ice effects) with the reconstructed electric field from tests in the anechoic chamber.}
\end{figure}

A typical electric field from the SPICE data is shown in Fig.~\ref{fig:efieldCompare}, overlaid with the corresponding electric field reconstructed from the anechoic chamber data. We observe that the IDL-1 pulser used in the 2019 anechoic chamber tests produced a lower amplitude than the 2018 SPICE data. This was confirmed in 2019, one month after the anechoic chamber measurement, when the same IDL-1 pulser was lowered into the SPICE hole. The resulting events recorded with station 51 were all consistently lower in amplitude than in the 2018 test. Therefore, we overlay a 2019 SPICE reconstructed electric field (which includes ice effects) with the reconstructed electric field obtained in the anechoic chamber. The SPICE electric fields appear identical between the 2018 and 2019 setup, modulo an overall scaling in amplitude. As seen in Fig.~\ref{fig:efieldCompare}, the main pulse of the electric fields between the SPICE hole data and the anechoic chamber data is similar in frequency and amplitude, which demonstrates that the applied ice corrections (frequency-dependent ice attenuation and bending of the signal) are well-understood. There is evidence of interference in both measurements, but the two setups have different geometries. Also, the frequency scaling of the anechoic data from in-air to in-ice is only a first order approximation, and the dipole emitter might behave differently when placed in ice which can cause some of the residual differences. 

\begin{figure}[t]
\centering
\includegraphics[width=0.8\textwidth]{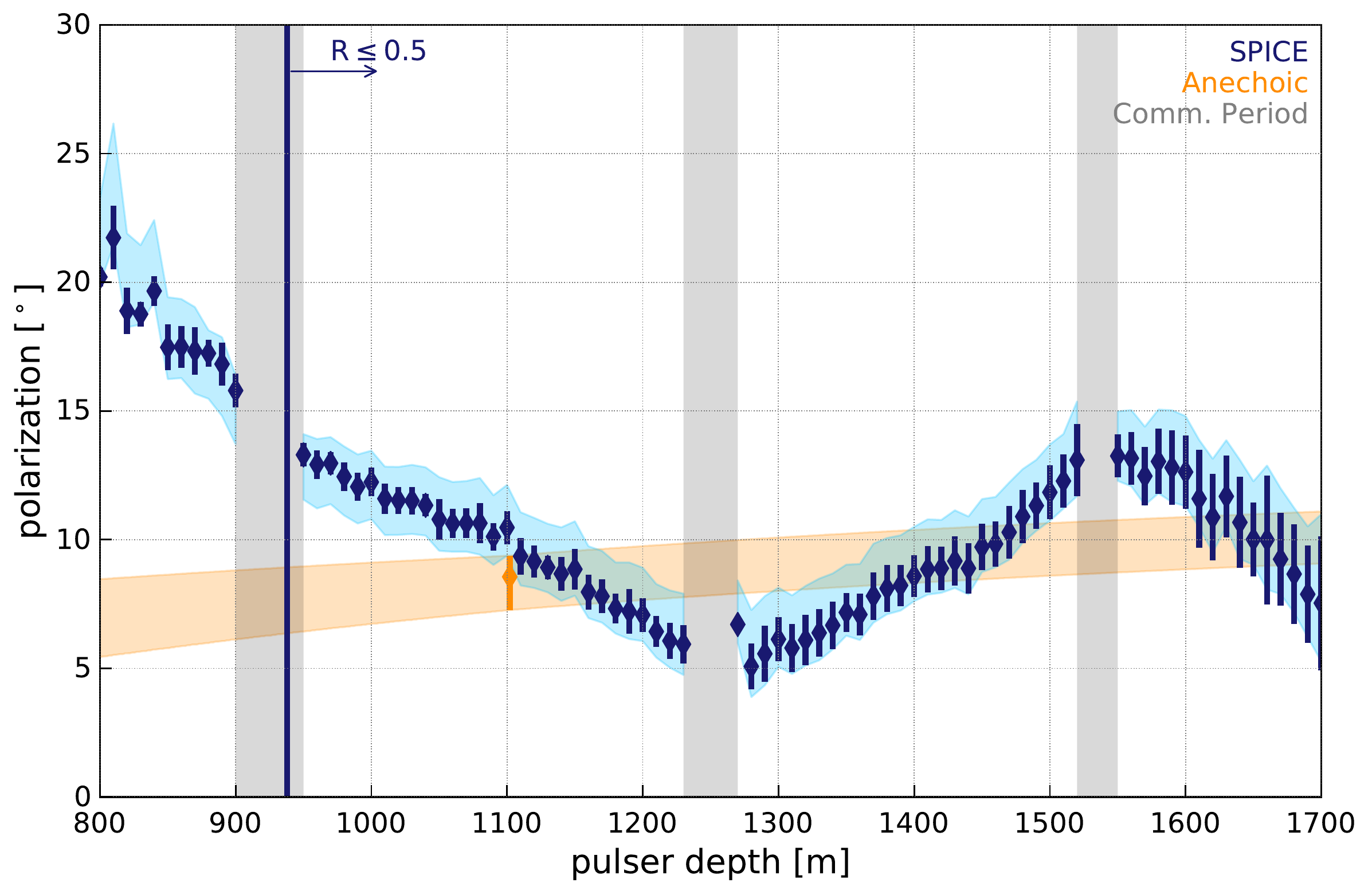}
\caption{\label{fig:polCompare} Measured polarization angle (blue data points) from 2018 SPICE hole experiment compared to measured polarization angle from tests in the anechoic chamber (orange band). The vertical blue line at \SI{938}{m} indicates the boundary for which the reflection coefficient is 0.5. The gray bands shows the periods where the station was in communication mode and thus not taking data. The SPICE data was averaged over \SI{10}{m} depths, and the 1$\sigma$ spread of the distribution averaged is shown with the blue error bars. The light blue shading indicates the systematic uncertainty on the reconstruction stemming from systematic uncertainties in the ARIANNA LPDA orientations. There is only one anechoic data point that fits in the depth ranges of the SPICE data and is indicated as an orange diamond; the error bar represents the spread of the 10 event average. The orange band shows the linear interpolation to the next data points, outside of the depth range plotted. For the
anechoic data the representative depth was calculated from the launch angle as in Fig.~\ref{fig:rays}.}
\end{figure}

\begin{figure}[t]
\centering
\includegraphics[width=0.6\textwidth]{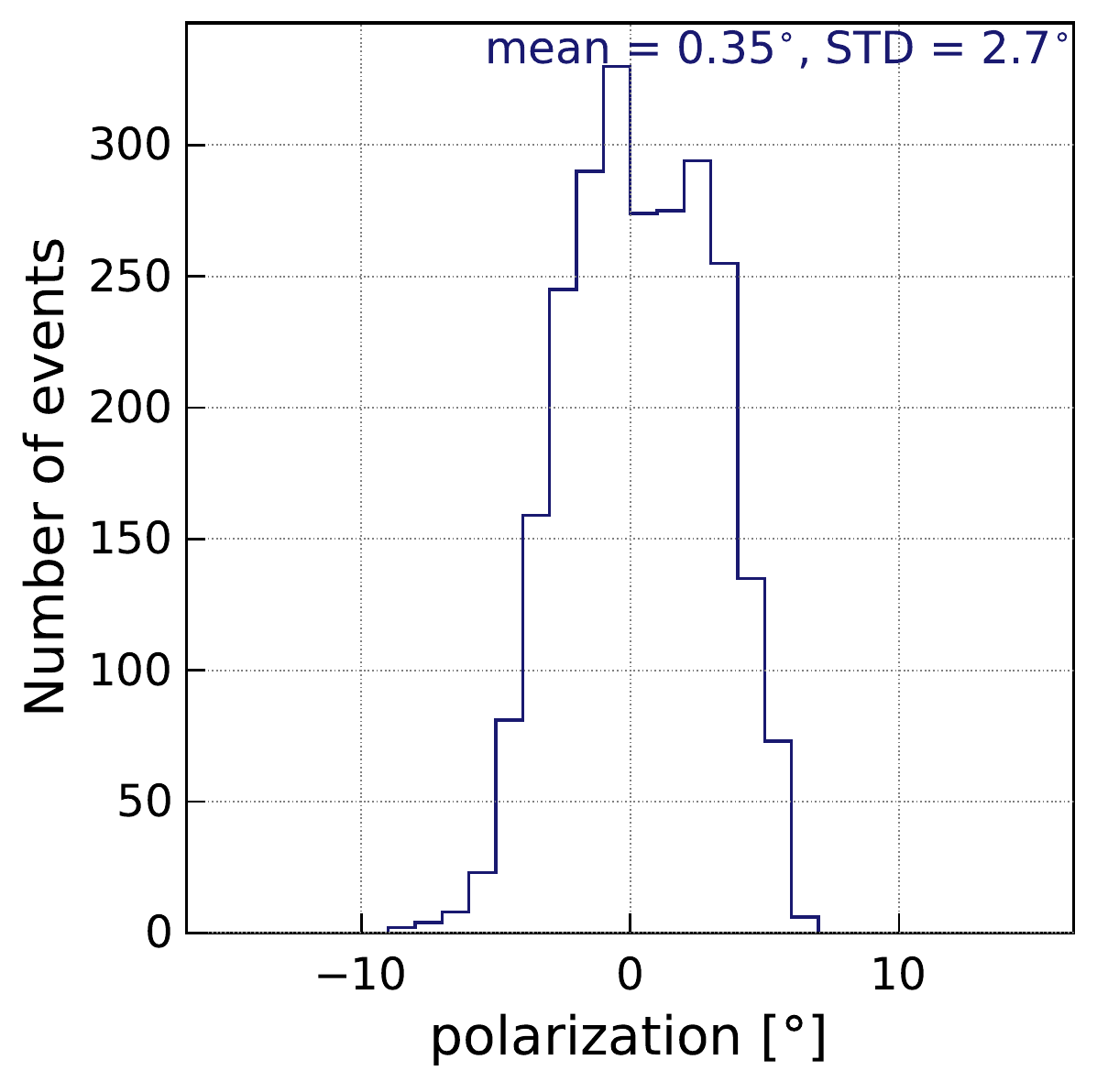}
\caption{\label{fig:histPol} Difference between measured polarization from 2018 SPICE hole experiment (without averaging) and measured polarization from tests in the anechoic chamber.}
\end{figure}

In Fig.~\ref{fig:polCompare},
we compare the reconstructed polarization from the SPICE data to the prediction from the anechoic chamber measurement (cf. Fig.~\ref{fig:expPol}), where the launch angle has been converted to depth according to Fig.~\ref{fig:rays}. The resulting polarization measurements are then averaged over \SI{10}{m} depths which results in roughly 30 polarization measurements being averaged together. This is shown as dark blue circles in Fig.~\ref{fig:polCompare}, where the error bars represent the 1$\sigma$ spread of the distribution averaged. The light blue shading in Fig.~\ref{fig:polCompare} represents systematic uncertainties of the measurement resulting from uncertainties in the orientation of the LPDA antennas. When comparing the SPICE measurements to the anechoic measurements, we exclude data where the reflection coefficient is greater than 0.5 as indicated by the vertical blue line in Fig.~\ref{fig:polCompare} just as we had done for the angular reconstruction of the LPDAs. The SPICE data reconstructs a polarization that scatters around \SI{9}{\degree}, whereas the anechoic data reconstructs the polarization at \SI{8}{\degree} - \SI{10}{\degree}. The histogram of Fig.~\ref{fig:histPol} shows the difference between SPICE measurement (on an event-by-event basis, i.e. without averaging) and the anechoic chamber prediction. We find a small mean offset of \SI{0.35}{\degree} and a scatter of \SI{2.7}{\degree}. We infer that we can make a precise polarization measurement for neutrino-induced Askaryan signals from the ability to determine the polarization of the radio pulser events.

There is some depth dependence seen in 
Fig~\ref{fig:polCompare}. In particular, the reconstructed polarization from the SPICE data oscillates around the prediction from the anechoic chamber measurement. We observe that the amplitude of the $\theta$ component decreases monotonically with depth, as expected from ice attenuation and 1/r field diminution. The $\phi$ component, which has a lower signal-to-noise ratio\footnote{using the standard definition of maximum signal amplitude divided by the RMS noise} ($\sim$4-8) than the $\theta$ component's signal-to-noise ratio ($\sim$20-40), also shows this trend but with an additional oscillation of its amplitude. This results in the observed oscillation in the polarization, which is itself defined as the ratio of the amplitudes of the two components (cf. Eq.~\eqref{eq:1} and \eqref{eq:2}).
Although the exact reason for this effect is not known, we have considered the following potential sources:
\begin{itemize}
    \item \textbf{Arrival Direction:} An uncertainty in the signal arrival direction will affect the antenna response pattern which is used in the polarization reconstruction. However, small angular changes of a few degrees have little impact on the antenna response. We find that changing the incident direction by $\pm$\SI{2}{\degree} does not change the oscillatory behavior seen in Fig.~\ref{fig:polCompare} and only leads to a depth-independent shift in the reconstructed polarization of $\pm$\SI{1}{\degree}.
    \item \textbf{Antenna response:} Boundary effects are hard to accurately model for antennas very close to a boundary. Since the receiver antennas are so shallow, the close proximity to the ice/air interface is likely to influence the antenna response. We have repeated the polarization reconstruction with antenna response patterns simulated for the LPDA immersed in finite firn (our nominal results), \SI{1}{m} and \SI{1}{cm} below the snow surface and did not observe any significant differences for transmitter depths below \SI{1}{km} where surface reflections are small. However, the different LPDAs might be impacted differently by the boundary due to small differences in the geometry or snow surface which could impact the reconstructed polarization. \footnote{We note that the antenna-related uncertainties will improve in a future ARIANNA-200 detector \cite{ARIANNA200} by installing the antennas deeper into the ice.} 
    \item \textbf{Ice profile:} If propagation through the ice affected the polarization, a monotonic increase or decrease of the polarization with emitter depth would be expected. Thus, attributing the oscillatory behaviour to ice properties is challenging and would require different inhomogenities for different paths. A prior analysis \cite{Barwick2018} demonstrated that local ice density fluctuations, particularly in the firn, can result in classically unexpected signal propagation modes. Without additional {\it in situ} studies, we cannot rule out the possibility that such effects contribute to our observations.
    \item \textbf{Change in emitter response with time:} If the emitted signal changed with time, then the polarization would also change with time and therefor depth. However, this is unlikely since ARIANNA observes the same polarization trend when analyzing SPICE data taken while the pulser was being lowered versus being raised.
    \item \textbf{Emitter characteristics:} If the emitted signal had some depth dependence, then the polarization would also change with depth. This might originate from depth dependent properties of the SPICE borehole, such as slight changes in the SPICE hole radius. Also, the emitter was lowered by a metal cable that will impact the response pattern of the emitting antenna, especially for launch angles close to the vertical. This may also result in the observed oscillatory behavior.
    \item \textbf{Reference measurement:} The anechoic chamber measurement was performed at discrete launch angles (cf. Fig.~\ref{fig:expPol}); only one laboratory launch angle lies within the corresponding range of emitter depths analyzed here. This reference point is shown as the orange diamond in Fig.~\ref{fig:polCompare}. The predicted polarization is obtained via linear interpolation to reference measurements corresponding to depths outside our depth range. Interestingly, the reconstructed polarization from the SPICE data matches the anechoic measurement at the \SI{1100}{m} reference point. A possible origin of the oscillatory behaviour is thus a change in the emitter characteristics with launch angle that was not captured by the discrete measurements performed in the anecoic chamber. 
\end{itemize}

From the discussion above and because the change in polarization originates from amplitude variations of the small $\phi$ polarization component, we speculate that a change of the emitter characteristics is the most likely origin of this effect. The $\phi$ polarization corresponds to the cross-polarization component, for which an ideal dipole should have zero transmission. Thus, a change of the cross-polarization amplitude with depth and launch angle seems plausible (cf section \ref{sec:bire}). This would also mean that the polarization can be measured much better because the scatter of \SI{2.7}{\degree} is largely determined by the oscillations. The scatter of the reconstructed polarization within a narrow depth range is often smaller than \SI{1}{\degree}. We also note the expected radio pulses from neutrinos will be cleaner: The signal will be the same in both polarization components in that the frequency spectrum and time domain behavior will be identical and only differ in amplitude which will facilitate the polarization reconstruction. This is in contrast to the SPICE transmitter which does not have the same frequency spectrum and time domain behavior in both polarization components.

\subsection{Polarization-dependent birefringence}
\label{sec:bire}
In birefringent media, signal propagation wave speeds are anisotropic \cite{hargreaves_1978}. For polar ice, which is known to be birefringent over a frequency range stretching from ultra-violet through radio, vertical gravitational pressure and lateral ice flow, at a given depth, break spatial symmetry. Although a perfectly hexagonal ice crystal has no preferred planar symmetry axis, the vertical strain profile of the ice sheet, physically originating in the hydrostatic pressure gradient towards lower surface elevations combined with `grounding' of the ice sheet on the bedrock below, results in a bulk distortion of the crystal fabric along the strain direction. At South Pole, for example, the surface ice flow velocity is approximately \SI{9}{m/yr}, decreasing to near zero at the bedrock (save for episodic `slip' events). Correspondingly, it is reasonable to expect that the birefringent axes align along i) the vertical, ii) the ice flow direction in the horizontal plane, and iii) a perpendicular to ice flow in the horizontal plane, with refractive indices $n_3$, $n_1$ and $n_2$, respectively (cf.~\cite{Jordan2020}). Numerically, below the firn layer, at depths (greater than 150 m) for which the ice has reached its asymptotic density, $n_2\approx n_3$, whereas $(n_2-n_1)/(n_2+n_1)\sim$0.1\%. Signals emitted from the SPICE core transmitter project into the three orthogonal planes (aka 'rotation planes', since a polarized continuous wave signal would appear to rotate as it advances owing to birefringent asymmetries) defined by these three axes at the source point. Each projected electric-field polarization vector within a given plane then has two components which independently propagate with wave velocities corresponding to the respective refractive indices for the axes defining that plane. The laboratory-measured values for these three refractive indices, coupled with the measured alignment of typical ice crystals (stacked vertically and elongated in the direction of ice flow) derived from SPICE core data ~\cite{SPICEbirefringenceData} can then be used to make absolute predictions for the relative time delays for the three components at the ARIANNA antenna measurement point. Accounting for the known directivity of the ARIANNA LPDAs or dipoles yields the expected amplitude for a given component, arriving with that time delay.

A vector from the SPICE core to the ARIANNA station is nearly coincident (cf.~Fig.~\ref{fig:stn51}) with the local ice-flow direction, simplifying the problem considerably. Anechoic chamber data indicates that the source transmitter dipole has a measured cross-polarization (corresponding to the $\vec{e}_\phi$ direction in spherical coordinates) amplitude, in the plane perpendicular to the propagation direction and the long axis of the antenna, of order 10\%. The dominant transmitter component is polarized perpendicular to the signal propagation direction and in the plane containing the dipole long axis (corresponding to the $\vec{e}_\theta$ direction in spherical coordinates). This component therefore projects into the $n_1-n_3$ and $n_2-n_3$ planes, but not $n_1-n_2$. At the receiver, the cross-polarization $\phi$ component parallel to $n_2$ is nearly simultaneous with the arrival of the vertically polarized signal (i.e. the $\theta$ component projected onto the vertical axis) since $n_2\approx n_3$. The horizontal component parallel to $n_1$ (i.e. the $\theta$ component projected onto the $n_1$ axis) leads by ${\cal O}$(10-20 ns) relative to the vertically-polarized signal, with an amplitude increasing with transmitter depth (cf.~\cite{Jordan2020}). 

In the following, we compare the pulse arrival time of the vertical and horizontal component parallel to the ice flow. These two components have both similar amplitude and a significant time delay is expected which can be compared to the measured signals.

\begin{figure}[t]
\centering
\includegraphics[width=0.8\textwidth]{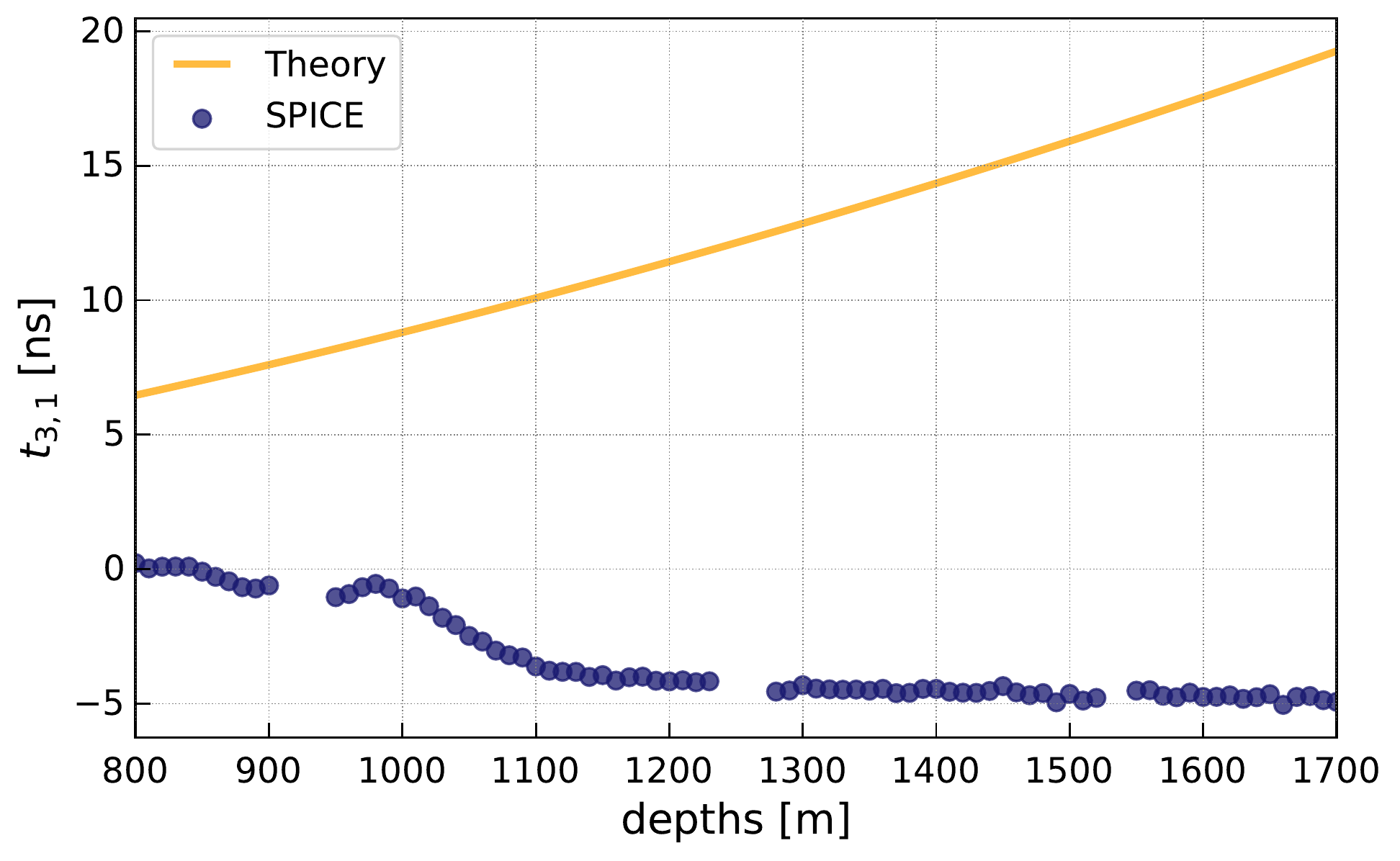}
\caption{\label{fig:bire}Distribution for $t_{3,1}$, defined as the signal arrival time of the dipoles minus the signal arrival time of the $\theta$ component of the LPDAs. Blue circles show the SPICE measured values as a function of depth. The theoretical prediction (orange line) is based on the model developed in \cite{Jordan2020} (see text for details).}
\end{figure}

The dipole antennas are only sensitive to vertically polarized signals. Thus, we use the electric field reconstructed from the dipole measurement to estimate the signal arrival time of the vertical component. The LPDAs are only sensitive to horizontally polarized signals and for our geometry, the $\theta$ polarization component projected onto the horizontal plane is approximately parallel to the ice flow. Thus, we use the $\theta$ component of the electric field derived from the LPDAs to estimate the signal arrival time of the horizontal component parallel to the ice flow. For both components we define the maximum of the Hilbert envelope as the signal arrival time. We denote $t_{3,1}$ as the signal arrival time of the dipoles minus the signal arrival time of the $\theta$ component of the LPDAs. The measured $t_{3,1}$ distribution is presented in Fig.~\ref{fig:bire}, together with the theoretical prediction described above (cf.~\cite{Jordan2020}).

Birefringence would appear as a depth-dependent time delay between the signal arrival for the two antenna types. However, we observe an approximately constant time delay of less than \SI{5}{ns}. From \SI{1000}{m} to \SI{1150}{m} the time delay changes from $\approx$\SI{0}{ns} to \SI{4}{ns} which corresponds to the depth region where the reflection coefficient transitions from total internal reflection to negligible reflection. Hence, this small change in time delay might be attributed to a change in the interference with the reflected signal. In any case, the measurement does not show the nearly linear dependence with depth as predicted from our model and the absolute time delays are much smaller than the predicted values, which is consistent with previous measurements made with RICE data \cite{besson2010radio}. We therefore conclude that we do not observe evidence of birefringence for this geometry. 

The ARA experiment measured birefringent asymmetries derived from the same pulser data set described herein \cite{Allison2019a}; in contrast to ARIANNA, the signal propagation geometry for ARA is predominantly horizontal rather than vertical. ARA measures 20-30 ns signal arrival time differences for horizontally vs.~vertically polarized signals consistent, to within 30\%, of expectation based on the same $n_1$, $n_2$ and $n_3$ data used as input for the model/data comparison shown in Fig.~\ref{fig:bire}. Given that the SPICE ice fabric measurements do not explicitly measure the orientation of the ice crystals in the horizontal plane, one may consider whether relaxing the assumption that $n_1$ aligns with ice flow can alleviate this tension.
Within the context of our model, however, a simple azimuthal rotation of the underlying birefringent principal axes cannot reconcile the discrepancy between the ARA and ARIANNA measurements, which is further constrained by the nearly perfect alignment of the ARIANNA receiver station with the ice flow direction; a more refined ice model is therefore likely required.

\section{Discussion and Conclusion}
We presented the measurement of calibration pulser signals, which were emitted deep in the ice at South Pole, with LPDA and dipole antennas placed slightly below the surface. The variable depth of the emitter and the large propagation distances of up to \SI{2}{km} validated the modeling of the signal propagation with high precision. 

We measured the signal arrival directions and compared it with the expectation which was computed from the emitter depth and a detailed calculation of the bending of the signal trajectories while propagating through the firn. We observe a negligible offset between measurement and prediction with an event-by-event scatter to better than \SI{0.4}{\degree}. This result is of direct importance for the measurement of neutrinos: The effect of the ice on the propagation direction can be corrected with high precision which is important for reconstructing the neutrino direction. The corresponding uncertainty from ice modelling is likely much smaller than \SI{0.4}{\degree} as this scatter is mostly due to statistical event-by-event uncertainties. No evidence for a systematic shift in reconstructed direction with depth was found. 

We reconstructed the three-dimensional incident electric field using two pairs of orthogonal oriented LPDA antennas and compared it with a reference measurement of the emitter in an anechoic chamber. After correcting for detector response and ice attenuation, we find agreement in amplitude, pulse shape and frequency content. This shows that the attenuation of radio signals is well understood and that the propagation through the ice does not lead to any significant distortion of the radio pulse. 

We also calculated the polarization from the reconstructed electric fields. We find a good agreement with the reference measurement of the anechoic chamber with an offset of \SI{0.35}{\degree} averaged over all depths, and a scatter of \SI{2.7}{\degree}. We observe an oscillation of the reconstructed polarization with depth which is likely due to changes in the emitter characteristics which would suggest that the polarization can be measured with even higher precision. Further studies are needed to find the origin of this effect and are planned for the future.

We do not observe birefrigence effects in this particular station and transmitter geometry. This is different to what is observed at nearly horizontal propagation at greater depth. These results encourage the development of an improved model for the description of the birefringence of the South Polar ice.

These results are of direct importance for the reconstruction of the direction and energy of neutrinos. The neutrino direction is a function of a) the signal arrival direction corrected for bending in the firn, b) the polarization, and c) an additional weak dependence on the viewing angle. The resolution of the neutrino direction is approximately the square root of the quadratic sum of the individual uncertainties of the three parameters. This analysis showed that uncertainties in the ice modelling will affect the neutrino direction resolution by not more than \SI{3}{\degree} and likely less depending on the origin of the scatter in the polarization reconstruction. 

\section{Acknowledgements}
We thank Tom Jordan for valuable discussions of birefringence.

We are grateful to the U.S. National Science Foundation-Office of Polar Programs, the U.S. National Science Foundation-Physics Division (grant NSF-1607719) for granting the ARIANNA array at Moore's Bay. Without the invaluable support of the people at McMurdo, the ARIANNA stations would have never been built.

We acknowledge funding from the German research foundation (DFG) under grants GL 914/1-1 (CG) and NE 2031/2-1 (DGF, ANe, IP, CW), and the Taiwan Ministry of Science and Technology (JN, SHW).  HB acknowledges support from the Swedish Government strategic program Stand Up for Energy. DB and ANo acknowledge support from the MEPhI Academic Excellence Project (Contract No.  02.a03.21.0005) and the Megagrant 2013 program of Russia, via agreement 14.12.31.0006 from 24.06.2013.

\bibliographystyle{JHEP}
\bibliography{bib}
\end{document}